\def\th {\tilde h}
\def\overbar#1{{\overline{#1}}}
\def\hh {\hat h}
\def\half {{1\over 2}}
\def\be{\begin{equation}}
\def\ee{\end{equation}}
\def\ba{\begin{eqnarray}}
\def\ea{\end{eqnarray}}
\def\nen{\nonumber}
\def\Tau {{\cal T}}

\documentclass[aps,epsf,preprint]{revtex4}
\begin{document}
%\tightenlines

\title{False loss of coherence}
\author{W. G. Unruh}
\affiliation{ CIAR Cosmolgy Program, Dept. of Physics\\University of B.C.\\Vancouver, 
Canada V6T 1Z1\\
      ~    \\
email:unruh@physics.ubc.ca\\
~
Published in {\bf Relativistic
Quantum Measurement and Decoherence}, ed F. Petruccione , Springer Velag(2000)}

\begin{abstract}
The loss of coherence of a quantum system coupled to a heat bath as expressed 
by the reduced density matrix is shown to lead to the miss-characterization 
of some systems as being incoherent when they are not.  
The spin boson problem and the harmonic oscillator with massive scalar field 
heat baths are given as examples
 of reduced incoherent density matrices  which nevertheless
still represent perfectly coherent systems.
\end{abstract}

\maketitle

\section{Massive Field Heat Bath and a Two Level System}
How does an environment affect the quantum nature of a system? The standard 
technique is to look at the reduced density matrix, in which one has traced 
out the environment variables. If this changes from a pure state to  a 
mixed state ( entropy $Tr\rho \ln\rho$ not equal to zero) one argues that 
the system has lost quantum coherence, and quantum interference effects are suppressed
. However this criterion is too strong.
There are couplings to the environment which are such that this reduced density
matrix has a high entropy, while the system alone   retains virtually
all of its original quantum coherence certain experiments.

The key idea is that the external environment can be different for different 
states of the system. There is a strong correlation between the system and the environment.
As usual, such correlations  lead to decoherence in the reduced density matrix.
However, the environment in these cases is actually
tied to the system, and is adiabatically dragged along by the system. Thus 
although the state of the environment is different for the two states, 
one can manipulate the system alone so as to cause these apparently incoherent 
states to interfere with each other. One simply causes a sufficiently slow change in the
system so as to drag the environment variables into common states so the quantum 
interference of the system can again manifest itself.

An  example is if one looks at an electron with its attached electromagnetic 
field. Consider the electron at two different positions. The static coulomb 
field of the two charges differ, and thus the states of the electromagnetic 
field differ with the electron in the two positions. These differences 
can be sufficient to cause the reduced electron wave function loose
coherence  for a state which is a coherent sum of states located at these two
 positions. However, if one causes the system
to evolve so as to cause the electron in those two positions to come together
( eg, by having a force field such that the electron in both positions 
to be brought together at some central point for example), those two apparently 
incoherent states will interfere, demonstrating that the loss of coherence 
was not real.

Another example is light propagating through a slab of glass. If one simply
looks at the electromagnetic field, and traces out over the states of the atoms
in the glass, the light beams traveling through two separate regions of the
glass will clearly decohere-- the reduced density matrix for the electromagnetic
field will lose coherence in position space-- but those two beams of light will
also clearly interfere when they exit the glass or even when they are within the
glass.

The above is not to be taken as proof, but as a motivation for the further 
investigation of the problem. The primary example I will take will be of a
spin $\half$ particle (or other two level system). 
I will also examine a harmonic oscillator as the system of interest.
In both cases, the heat bath will be a massive one dimensional scalar field. 
This heat bath
is of the general Caldeira Leggett type \cite{caldera}( and in fact is entirely equivalent to 
that model in general). The mass of the scalar field will be taken to be larger than
the inverse time scale of the dynamical behaviour of the system. This is not to be taken
as an attempt to model some real heat bath, but to display the phenomenon in its clearest form.
Realistic heat baths will in general also have low frequency excitations which will introduce
other phenomena like damping and genuine loss of coherence into the problem.

\section{Spin-$\half$ system}

Let us take as our first example that of  a spin-$\half$ system coupled 
to an external environment. We will take this external environment 
to be a one dimensional massive scalar field. The coupling to the spin 
system will be via purely  the $3$ component of the spin. I will use the
velocity coupling which I have used elsewhere as a simple example of an 
environment (which for a massless field is completely equivalent to the
 Caldeira Leggett model). The Lagrangian is
\be
L=\int \half ((\dot \phi(x))^2 - (\phi(x)')^2+m^2\phi(x)^2 +2\epsilon \dot 
\phi(x)h(x)\sigma_3 ) dx
\ee
which gives the Hamiltonian
\be
H=\int \half((\pi(x)-\epsilon h(x)\sigma_3)^2+(\phi(x)')^2+m^2\phi(x)^2) dx
\ee
$h(x)$ is the interaction range function, and its Fourier transform is related  to 
the spectral response function of Leggett and Caldeira.

This  system is easily solvable. I will look at this system in the following 
way. Start initially with 
the field in its free ($\epsilon=0$) vacuum state, and the system is in the +1
eigenstate of $\sigma_1$. I will start with the coupling $\epsilon$ initially zero and 
gradually increase it to some large value.  I will look at the reduced density matrix for
the system, and show that   it  reduces one which is almost the
 identity matrix ( the maximally incoherent density
matrix) for strong coupling. Now I let $\epsilon$ slowly drop to zero again.
 At the end of the procedure,
the state of the system will again be found to be in the original eigenstate 
of $\sigma_1$. The intermediate  maximally incoherent density matrix would seem to
imply that the system no longer has any quantum coherence. However this lack of
coherence is illusionary. Slowly decoupling the system from the environment 
should in the usual course simply maintain the incoherence of the system 
Yet here, as if by magic, an almost completely incoherent density matrix
magically becomes coherent when the system is decoupled from the environment. 

In analyzing the system, I will look at the states of the field corresponding to the two possible
$\sigma_3$ eigenstates of the system. These two states of the field are 
  almost orthogonal for strong coupling. However they correspond to fields tightly
bound to the spin system. As the coupling is reduced, the two states of
the field adiabatically come closer and closer together until finally they coincide when
$\epsilon $ is again zero. The two states of the environment are now the same, 
there is no correlation between the environment and the system,
and the system regains its coherence.

The density matrix for the spin system can always be written as 
\be
\rho(t) = \half(1+\vec\rho(t)\cdot\vec\sigma)
\ee
where 
\be
\vec\rho(t)= Tr(\vec\sigma \rho(t))
\ee
We have
\be
\vec\rho(t) = Tr\left(\vec\sigma \Tau[e^{-i\int_0^t H dt}]
\half(1+\vec\rho(0)\cdot\vec\sigma)R_0 
\Tau[e^{-i\int H dt}]^\dagger \right)
\ee
where $R_0$ is the initial density matrix for the field (assumed to be the vacuum),
and $\Tau[]$ is the time ordering operator. (Because $\epsilon$ and thus $H$ 
is time dependent, the $H$ at different times do not commute. this leads to 
requirement for the time ordering in the expression. As usual, the time 
ordered integral is the way of writing the time ordered product
 $\prod_n e^{-iH(t_n) dt}= e^{-iH(t)dt} e^{-iH(t-dt) dt}....e^{-iH(0)dt}$.)

Let us first calculate $\rho_3(t)$. We have
\ba
\rho_3(t)&= Tr \left(\sigma_3 \Tau[e^{-i\int_0^t H dt}]
\half(1+\vec\rho(0)\cdot\vec\sigma)R_0 
\Tau[e^{-i\int H dt}]^\dagger \right)
\\
&=Tr \left(\Tau[e^{-i\int_0^t H dt}]\sigma_3
\half(1+\vec\rho(0)\cdot\vec\sigma)R_0 
\Tau[e^{-i\int H dt}]^\dagger \right)
\\
&=Tr\left(\sigma_3\half(1+\vec\rho(0)\cdot\vec\sigma)R_0  \right)
\\
&= \rho_3(0)
\ea
because $\sigma_3$ commutes with $H(t)$ for all $t$.
We now define
\be
\sigma_+=\half(\sigma_1 +i\sigma_2)=|+><-|;~~~~~~\sigma_-=\sigma_+^\dagger
\ee
Using $\sigma_+\sigma_3 = -\sigma_+$ and $\sigma_3\sigma_+=\sigma_+$
we have
\ba
Tr &~& \left(\sigma_+ \Tau[e^{-i\int_0^t H dt}]
\half(1+\vec\rho(0)\cdot\vec\sigma)R_0 
\Tau[e^{-i\int H dt}]^\dagger \right)
\nen \\
&= &Tr_\phi \left(\Tau[e^{-i\int (H_0-\epsilon(t)\int\pi(x) h(x) dx)dt}]^\dagger\right.
\\
&~&~~~~~~~~~~~~~\left. \Tau[e^{-i\int ( H_0+ \epsilon(t)\int\pi(x)h(x) dx) dt}]\right) <-|
 \half(1+\vec\rho(0)\cdot\vec\sigma)|+>
\nen\\
&= (\rho_1(0) +i\rho_2(0) ) J(t)
\nen
\ea
where $H_0$ is the Hamiltonian with $\epsilon=0$, i.e., the free Hamiltonian
for the massless scalar field and 
\be
J(t)=Tr_\phi \left( \Tau [ e^{-i\int(H_0 -\epsilon(t)\int\pi(x)h(x) dx)dt} 
]^\dagger \Tau[e^{-i\int(H_0 +\epsilon(t)\int\pi(x)h(x) dx)dt}]R_0 \right)
 \ee
Breaking up the time ordered product in the standard way into a large number 
of small time steps,
using the fact that $e^{-i\epsilon(t) \int h(x)\phi(x) dx}$ is the displacement
operator for the field momentum through a distance of $\epsilon(t)h(x)$, 
and commuting  the free field Hamiltonian terms through, this can be written
as
\ba
J(t)&=& Tr_\phi \left( e^{-i\epsilon(0)\Phi(0)}
\prod_{n=1}^{t/dt}\left[ e^{-i(\epsilon(t_n)-\epsilon(t_{n-1})\Phi(t_n)}\right]  \right.
\nen \\
    &~&~~~~~\left. e^{i\epsilon(t)\Phi(t)} e^{i\epsilon(t)\Phi(t)} \prod_{n=t/dt}^{1}\left[
e^{i\epsilon(t_n-\epsilon(t_{n-1}))\Phi(t_n)}\right] e^{i\epsilon(0)\Phi(0)} 
R_0 \right)
\ea
where $t_n = ndt$ and dt is a very small value, $\Phi(t)=\int h(x)\phi(t,x) dx$ 
and $\phi_0(t,x)$ is the free field Heisenberg field operator.  
Using the Campbell-Baker-Hausdorff formula, realizing that the commutators 
of the $\Phi$s are c-numbers, and noticing that these c-numbers cancel 
between the two products, we finally get
\ba
J(t)= Tr_\phi\left( e^{2i(\epsilon(t)\Phi(t)-\epsilon(0)\Phi(0) +\int_0^t 
\dot\epsilon(t')\Phi(t') dt')} R_0\right)
\ea
from which we  get
\ba
ln(J(t)) = - 2 Tr_\phi\left( R_0 \left(\epsilon(t)\Phi(t)-\epsilon(0)\Phi(0) 
+\int_0^t \dot\epsilon(t')\Phi(t') dt'\right)^2 \right)
\ea
I will assume that $\epsilon(0)=0$, and that $\dot\epsilon(t)$ is very 
small, and that it can be neglected. ( The neglected terms 
are of the form 
$$\int\int\dot\epsilon^2<\Phi(t')\Phi(t")>dt'dt" 
\approx \dot\epsilon^2 t \tau <\Phi(0)^2>$$
which for a massive scalar field has $\tau$, the coherence time scale, 
$\approx 1/m$. Thus,  as we let $\dot\epsilon$ go to zero   these terms
go to zero.)

We finally have
\ba
ln(J(t))& =& -2\epsilon(t)^2 <\Phi(t)^2>
\nen\\
&=& -2\epsilon(t)^2 \int |\hat h(k)|^2{1\over \sqrt(k^2+m^2)} dk
\ea
Choosing $\hh(k)=e^{-\Gamma |k|/2}$, we finally get
\be
ln(J(t))=- 4\int_0^\infty\epsilon(t)^2{ e^{-\Gamma |k|} dk \over sqrt(k^2+m^2)}
\ee
This goes roughly as $\ln(\Gamma m) $ for small $\Gamma m$,
(which I will assume is true). For $\Gamma$ sufficiently small, this  makes 
$J$ very small, and the density matrix reduces to essentially diagonal 
form ($\rho_z(t)\approx\rho_y(t)\approx 0$, $\rho_z(t)=\rho_z(0)$.)

However it is clear that if $\epsilon(t)$ is now lowered slowly to zero, 
the decoherence factor $J$ goes back to unity, since it depends only
 on $\epsilon(t)$. The density matrix  now has
exactly its initial form again. The loss of coherence at the intermediate times
was illusionary. By decoupling the system from the environment after the 
coherence had been lost, the coherence is restore. this is in contrast
 with the naive expectation in which 
the loss of coherence comes about because of the correlations between the 
system and the environment. Decoupling the system from the environment 
should not in itself destroy that correlation, and should not reestablish 
the coherence.

The above approach, while giving the correct results, is not very transparent 
in explaining what is happening. Let us therefor take a different approach. 
Let us solve the Heisenberg equations of motion for the field $\phi(t,x)$. 
The equations are ( after eliminating $\pi$) 
\ba
\partial_t^2 \phi(t,x) - \partial_x^2 \phi(t,x) +m^2 \phi(t,x) =-\dot \epsilon(t) 
\sigma_3 h(x)
\\
\pi(t,x)= \dot\phi(t,x) +\epsilon(t) h(x) \sigma_3
\ea
If $\epsilon$ is slowly varying in time, we can solve this approximately 
by
\ba
\phi(t,x)= \phi_0(t,x) + \dot \epsilon(t)\int {1\over 2m}e^{-m|x-x'|} h(x')dx' 
\sigma_3 +\psi(t,x)\epsilon(0)\sigma_3
\\
\pi(t,x)= \dot \phi_0(t,x) + \epsilon(t) h(x) \sigma_3 +\dot\psi(t,x)\epsilon(0)\sigma_3
\ea
 where $\phi_0(t,x)$ and $\pi_0(t,x)$ are  free field solution  to the 
equations of motion in 
absence of the coupling, with the same initial conditions 
 \ba
\dot\phi_0(0,x)=\pi(0,x)
\\
\phi_0(0,x)=\phi(0,x)
\ea
, while $\psi$ is also a solution of the free field equations but with
initial conditions \ba
\psi(0,x)=0
\\
\dot \psi(0,x)= -h(x).
\ea

 If we examine this for the two possible eigenstates of $\sigma_3$, we 
find the two solutions
\ba
\phi_{\pm}(t,x)\approx  \phi_0(t,x) \pm (\dot\epsilon(t)\int{1\over 2m} 
e^{-m|x-x'|} h(x')dx' +\psi(t,x))
\\
\pi_\pm(t,x)\approx \dot \phi_0(t,x) + O(\dot\epsilon)
 \pm   (\epsilon(t) h(x) +\epsilon(0) \dot\psi(t,x) )
 \ea
These solutions neglect terms of higher derivatives in $\epsilon$.
The state of the field is the vacuum state of $\phi_0,\pi_0$. $\phi_\pm $ 
and $\pi_\pm$ are equal to this initial field plus c number fields. Thus 
in terms of the $\phi_\pm $ and $\pi_\pm$, the state is a coherent state 
with non-trivial displacement from the vacuum.   Writing the fields in terms 
of their creation and annihilation operators,
\ba
\phi_{\pm}(t,x) = \int A_{k\pm}(t) e^{ikx} + A^\dagger_{k\pm} e^{-ikx}{ 
dk
\over \sqrt{2\pi\omega_k}}\\
\pi_\pm (t,x)= i\int A_{k\pm}(t) e^{ikx} 
- A^\dagger_{k\pm} e^{-ikx} \sqrt{k^2+m^2\over
2 \pi} dk
\ea

we find that we can write $A_{k\pm}$ in terms of the initial operators 
$A_{k0}$
as
\be
A_{k\pm}(t)\approx A_{k0}e^{-i\omega_k t}\pm \half i(\epsilon(t)- \epsilon(0) 
e^{-i\omega_k t}) (h(k)/\sqrt{ \omega_k}  + O(\dot\epsilon(t))  )
\ee
where $\omega_k=\sqrt{k^2+m^2}$. Again I will neglect the terms of order $\dot \epsilon$ 
 in comparison with the $\epsilon$ terms.
Since the  state is the vacuum state with respect to the initial 
operators $A_{k0}$, it will be a coherent state with respect to the operators 
$A_{k\pm}$, the  annihilation operators for the field at time $t$.
We thus have two possible coherent states for the field, depending on whether 
the spin is in the upper or lower eigenstate of $/sigma_3$. But these two 
coherent states will have a small overlap.  
If $A|\alpha>=\alpha |\alpha>$ then we have
\be
|\alpha> = e^{\alpha A^\dagger-|\alpha|^2/2}|0>
\ee
Furthermore, if we have two coherent states $|\alpha> $ and $|\alpha'>$, then
the overlap is given by
\be
<\alpha|\alpha'> = <0|e^{\alpha^* A-|\alpha|^2/2} e^{\beta A^\dagger
-|\beta|^2/2}|0>= e^{\alpha^*\beta-(|\alpha|^2+|\beta|^2)/2}
\ee
In our case, taking the two states $|\pm_\phi>$, these correspond to coherent states with 
\be
\alpha=-\alpha'= \half i(\epsilon(t)- \epsilon(0) e^{-i\omega_k t}) 
= \half  i \epsilon(t)
 h(k)/\sqrt{ \omega_k}
\ee
Thus we have
 \be
<+_\phi,t|-_\phi,t> = \prod_k e^{-\epsilon(t)^2 |h(k)|^2/(k^2+m^2)} 
= e^{-\epsilon(t)^2\int {|h(k)|^2 \over \omega_k}dk}= J(t).
\ee
Let us assume that we began with the state of the spin as ${1\over \sqrt{2}} (|+>+|->)$.
The state of the system at time t in the Schroedinger representation is
${1\over \sqrt{2}} \left( |+>|+_\phi(t)>+|->|-_\phi>\right)$
and the reduced density matrix is
\be
\rho = \half (|+><+|+|-><-| + J^*(t) |+><-| +J(t) |-><+|).
\ee
The off diagonal terms of the density matrix are suppressed by 
the function $J(t)$. $J(t)$ however depends only on $\epsilon(t)$ and thus
, as long as we keep $\dot\epsilon$ small, the loss of coherence represented by
$J$ can be reversed simply by  decoupling the system from the environment slowly.

The apparent decoherence comes about precisely because the 
system in either the two eigenstates of $\sigma_3$ drives the field into two different
coherent states. For large $\epsilon$, these two states have small overlap. 
However, this distortion of the state of the field is tied to the system. 
$\pi$ changes only locally, and the changes in the field caused by the system
do not radiate away.   As $\epsilon$
slowly changes, this bound state of the field also slowly changes in concert . 
 However if one examines only the system, one
sees a loss of coherence because the field states have only a small overlap
 with each other.

The behaviour is very different if the system or the interaction changes 
rapidly. In that case the decoherence can become real. As an example, consider 
the above case in which $\epsilon(t) $ suddenly is reduced to zero. In 
that case, the field is left as a free field, but a free field whose state 
( the coherent state) depends on the state of the system. In this case 
the field radiates away as real ( not bound) excitations of the scalar 
field. The correlations with the system are carried away, and even if the 
coupling were again turned on, the loss of coherence would be permanent.

\section{Oscillator}

For the harmonic oscillator coupled to a heat bath, the Hamiltonian can 
be taken as
\be
H= \half\int (\pi(x)-\epsilon(t)q(t)\th(x))^2 + (\partial_x\phi(x))^2 +m^2\phi(t,x)^2dx  
+\half (p^2+\Omega^2q^2)
\ee

Let us  assume that  $m$ is much larger than $\Omega$ or that the
 inverse time rate of change of $\epsilon$. The 
solution  for
the field is given by 
\ba
\phi(t,x)\approx \phi_0(t,x) +\psi(t,x)\epsilon(0)q(0) 
- \dot\overbar{\epsilon(t)q(t)} \int {e^{-m|x-x'|}\over 2m}h(x')dx'
\\
\pi(t,x)\approx\dot\phi(t,x)+\dot\psi(t,x)\epsilon(0)q(0)
 - \ddot\overbar{\epsilon(t)q(t)} \int {e^{-m|x-x'|}\over 2m}h(x')dx'
+\epsilon(t)q(t)h(x)
\ea
where again $\phi_0$ is  the free field operator, $\psi$ is a free field 
solution with $\psi(0)=0$, $\dot\psi(0)= -h(x)$. Retaining terms only 
of the lowest order in $\epsilon$
  \ba
\psi(t,x)\approx\phi_0(t,x)
\\
\pi(t,x)\approx \dot\phi(t,x) +\epsilon(t) q(t) h(x)
\ea
The equation of motion for $q$ is
\ba
\dot q(t)= p(t)
\\
\dot p(t)= -\Omega^2 q +\epsilon(t)\dot \Phi(t)
 \ea
where $\Phi(t)= \int h(x)\phi(t,x)dx$. Substitution in the expression 
for $\phi$, we get
\be
\ddot q(t) +\Omega^2 q(t)\approx \epsilon\dot(\Phi_0(t)) +
  \epsilon(t)\ddot\overbar{\epsilon(t)q(t)} \int \int h(x)h(x'){e^{-m|x-x'|}\over 2m}dxdx'
\ee
Neglecting the derivatives of $\epsilon$ (i.e., assuming that 
$\epsilon$ changes slowly even on the time scale of $1/\Omega$), this becomes
\be
\left( 1+\epsilon(t)^2\int \int h(x)h(x'){e^{-m|x-x'|}\over 2m}dxdx'\right)\ddot q 
+\Omega^2 q = \partial_t(\epsilon(t)\Phi(t))
\ee
The interaction with the field thus renormalizes the mass of the oscillator to
$$M=\left( 1+\epsilon(t)^2\int \int h(x)h(x')\right)$$
The solution for $q$ is thus
\be
q(t)\approx q(0)\cos(\int_0^t \tilde\Omega(t)dt)
+{1\over \tilde\Omega} \sin(\int_0^t\tilde\Omega(t) dt ) p(0)
+{1\over \tilde\Omega}\int_0^t  \sin(\int_{t'}^t\tilde\Omega(t) dt )
\partial_t(\epsilon(t')\dot\overbar{\epsilon(t)\Phi_0(t')}dt'
\ee
where $\tilde\Omega(t) \approx \Omega/\sqrt{M(t)}$.

The important point is that the forcing term dependent on
 $\Phi_0$ is a rapidly oscillating term of frequency at least $m$.
 Thus if we look for example at $<q^2>$, the deviation from the free evolution
of the oscillator (with the renormalized mass) is of the order of
$\int \sin(\tilde\Omega t-t')\sin(\omega(t-t") <\dot\Phi_0(t')\dot\Phi_0(t")>dt'dt"$. But
$<\dot\Phi_0(t')\dot\Phi_0(t")>$ is a rapidly oscillating function of frequency at least
$m$, while the rest of the integrand is a slowly varying function with frequency much
less than $m$, Thus this integral will be very small ( at least $\tilde\Omega/m$ but typically
 much smaller than this depending on the time dependence of $\epsilon$). Thus the deviation
of $q(t)$ from the free motion will in general be very very small, 
and I will neglect it.

 Let us now look at the field. The field is put into a coherent state which depends 
on the value of $q$, because $\pi(t,x)\approx\dot\phi_0(t,x)+ \epsilon(t) q(t) h(x)$
Thus 
\be
A_k(t)\approx a_{0k} e^{-i\omega_k t} +i\half \hh(k) \epsilon(t)q(t)/\omega_k
\ee
The overlap integral for these coherent states with various values of $q$ is
\be
\prod_k<i\half \hh(k) \epsilon(t)q/\omega_k |i\half\hh(k) \epsilon(t)q'/\omega_k >
=e^{- {1\over 8}\int|\hh(k)|^2 dk(q-q')^2}
\ee
 The density matrix for the Harmonic oscillator is thus
\ba
\rho(q,q') =  \rho_0(t,q,q')e^{- {1\over 8}\int|\hh(k)|^2 dk(q-q')^2}
\ea
where $\rho_0$ is the density matrix for a free harmonic oscillator (with the renormalized mass).

Ie, we see a strong loss of coherence of the off diagonal terms of the density matrix. 
However this loss of coherence is false.  If we take the initial state for example
with two packets widely separated in space, these two packets will loose their coherence.
However, as time proceeds, the natural evolution of the Harmonic oscillator will bring
 those two packets together ($q-q'$ small across the wave packet). 
For the free evolution they would then interfere. They still
 do. The loss of coherence which was apparent when the two packets were widely separated 
disappears, and the two packets interfere just as if there were no coupling to the environment.
The effect of the particular environment used  is thus to renormalise the mass,
 and to make the density matrix appear to loose coherence.

\section{Spin Boson Problem}

Let us now complicate the spin problem in the first section by introducing into the 
system a free Hamiltonian 
for the spin as well as  the coupling to the environment. Following the example of the
spin boson problem, let me introduce a free Hamiltonian for the spin of the form
$\half \Omega \sigma_1$, whose effect is to rotate the $\sigma_3$ states
 (or to rotate the vector $\vec\rho$ in the $2-3$ plane with frequency $\Omega$.

The Hamiltonian now is
\be
H= \half \left( \int (\pi(t,x)-\epsilon(t) h(x)\sigma_3)^2 
+(\partial_x \phi(x))^2 +m^2\phi(t,x)^2 dx +\Omega\sigma_1 \right)
\ee
where again $\epsilon(t)$ is a slowly varying function of time.
We will solve this in the manner of the second part the first section.

If we let $\Omega$ be zero, then the eigenstates of $\sigma_z$ are
 eigenstates of the Hamiltonian. The field Hamiltonian
 ( for constant $\epsilon$) is given by
\be
H_\pm = \half \int (\pi - (\pm \epsilon(t)h(x)))^2 + (\partial_x\phi)^2 dx.
\ee
Defining $\tilde \pi= \pi - (\pm h(x))$, $\tilde\pi$ has the same commutation
 relations with $\pi $ and $\phi$ as does $\pi$. Thus in terms of $\tilde \pi$ 
we just have the Hamiltonian for the free scalar field. The instantaneous 
minimum energy state is therefor the ground state energy for the free 
scalar field for both $H_\pm$.
 Thus the two states are degenerate in energy. In terms of the operators $\pi$
and $\phi$, these ground states are coherent states with respect to 
the vacuum state of the original uncoupled 
($\epsilon=0$) free field, with the displacement of each mode given  by
\be
a_k |\pm> = \pm i\epsilon(t) {h(k)\over \sqrt{\omega_k}} |\pm>
\ee
or
\be
|\pm> = \prod_k | \pm \alpha_k> |\pm>_{\sigma_3}
 \ee
where the $|\alpha_k>$  are coherent states for the $k^{th}$ modes 
with coherence parameter $\alpha_k= i \epsilon(t){h(k)\over \sqrt{\omega_k}}$, and the 
states $|\pm>_{\sigma_3}$ are the two eigenstates  of $\sigma_3$. (In the following I
will eliminate the $\prod_k$ symbol.)
The energy to the next excited state in each case is just $m$, the mass of the 
free field.

We now introduce the $\Omega\sigma_x$ as a perturbation parameter. 
The two lowest states ( and in fact the excited states) are two fold degenerate.
Using degenerate perturbation theory to find the new lowest energy eigenstates, 
we must calculate the overlap integral 
of the perturbation between the original degenerate states   and
must then diagonalise the resultant matrix to lowest order in $\Omega$. 
The perturbation is $\half\Omega \sigma_1$ . All terms between the same
states are zero, because of the $<\pm|_{\sigma_3}\sigma_1|\pm >_{\sigma_3}=0$. 
Thus the only terms that survive for determining the lowest order
 correction to the lowest energy 
eigenvalues are 
\ba
\half<+|\Omega\sigma_1 |-> &=& \half<-|\Omega\sigma_1 |+>^*
\\
&=& \half\Omega\prod_k <\alpha_k|-\alpha_k>
=\half\Omega\prod_k e^{-2|\alpha_k|^2} 
\\
&=& \half\Omega e^{-2\int \epsilon(t)^2 |h(k)|^2/\omega_k dk} = \half \Omega J(t)
\ea
The eigenstates of energy thus have energy of  $E(t)_\pm=E_0 \pm \half \Omega J(t)$,
and the eigenstates are $\sqrt{\half}(|+>\pm|->)$ If epsilon varies slowly enough, 
the instantaneous energy eigenstates will be the actual adiabatic eigenstates 
at all times, 
and the time evolution of the system will just be in terms of these
 instantaneous energy eigenstates. Thus the system will evolve as
\ba
|\psi(t)> = &\sqrt{\half} e^{-iE_0 t}\left((c_++c_-) e^{-i\int \half\Omega_t J(t)dt} (|+>+|->) \right.
\\
&\left. ~~~~+ (c_- - c_+) e^{+i\int\half \Omega_t J(t)dt} (|+>-|->)\right)
\ea
where the $c_+$ and $c_-$ are the initial amplitudes for the 
$|+>_{\sigma_3}$ and $|->_{\sigma_3}$ states. 
The reduced density matrix for the spin system in the $\sigma_3$ 
basis can now be written as
\be
\vec\rho(t) = \left(J(t)\rho_{01}(t) ,J(t) \rho_{02}(t), \rho_{03}(t)\right)
\ee
where $\vec \rho_0(t) $ is the density matrix that one would obtain for a free 
spin half particle moving under the Hamiltonian 
$J(t)\Omega\sigma_1$.
\ba
\rho_{01}(t)=\rho_{1}(0)
\nen \\
\rho_{02}(t)=\rho_{2}(0) \cos(\Omega\int J(t')dt') +\rho_3(0)\sin(\Omega\int J(t') dt')
\\
\rho_{03}(t)=\rho_3(0) \cos(\Omega\int  J(t')dt') - \rho_2(0) \sin(\Omega \int J(t')dt
\nen
\ea
Thus if $J(t)$ is very small
 (ii.e., $\epsilon$ large) , we have a renormalized
 frequency for the spin system,  and the    the off diagonal terms
 (in the $\sigma_3$ representation) of the
density matrix are strongly suppressed by a factor of $J(t)$. 
Thus if we begin in an eigenstate of $\sigma_3$ the density 
matrix will begin with the vector $\vec\rho$ as a unit vector pointing 
in the 3 direction. As time goes on the 3 component gradually 
decreases to zero, but the 2 component increases only to the small
 value of $J(t)$. The system looks almost like a completely incoherent
 state, with almost the maximal entropy that the spin 
 system could have. However as we wait longer, the 3 component of 
the density vector reappears and grows back to its full unit value in the
 opposite direction, and
 the entropy drop to zero again. This cycle repeats itself endlessly 
 with the entropy oscillating between its minimum and maximum
value forever. 

The decoherence of the density matrix  ( the small off diagonal terms) 
obviously represent a false loss  of
coherence.  It represents a strong correlation between the system and the environment.
However the environment is bound to the system, and essentially forms
 a part of the system itself, at least  as long as the system moves slowly. However
   the reduced density matrix makes no distinction between
 whether or not the correlations between the system and 
the environment are in some sense bound to the system, or are
 correlations between the system and a freely propagating 
modes of the medium in which case the correlations can be extremely difficult to recover,
and certainly cannot be recovered purely by manipulations of the system alone.

\section{Instantaneous Change}

In the above I have assumed throughout that the system moves slowly 
with respect to the excitations of the heat bath.
Let us now look at what happens in the spin system if we rapidly
change the spin
of the system. In particular I will assume that the system is as in
 section 1,
a spin coupled only to the massive heat bath via the component
 $\sigma_3$ of the spin. Then at a time $t_0$, I instantly rotate the
 spin through some angle $\theta$ about the $1$ axis. In this case we will find that
the environment cannot adjust rapidly enough, and at least a part of the
loss of coherence becomes real, becomes unrecoverable purely through manipulations
of the spin alone.

The Hamiltonian is
\be
H= \half \int \left( (\pi(t,x)-\epsilon(t) h(x) \sigma_3)^2 
+(\partial_x\phi(t,x)^2 +m^2 \phi(t,x)\right)dx + \theta/2 \delta(t-t_0) \sigma_1
\ee
Until the time $t_0$ $\sigma_3 $ is a constant of the motion, and
 similarly afterward. Before the time $t_0$, the energy eigenstates
 state of the system are as in the last section given by
\be
|\pm,t> = \{|+>_{\sigma_3}|\alpha_k(t)> {\rm or }
\{|->_{\sigma_3}|-\alpha_k(t)>\}
\ee
An arbitrary state for the spin--environment system is given by
\be
|\psi> = c_+ |+> +c_- |->
\ee
Now, at time $t_0$, the rotation carries this to
\ba
|\phi(t_0)>= &c_+( \cos(\theta/2)|+>_{\sigma_3}+ i\sin(\theta/2)|->_{\sigma_3}|\alpha_k(t)>
\nen\\
&~~~+c_-( \cos(\theta/2)|->_{\sigma_3}
+ i\sin(\theta/2)|+>_{\sigma_3})|-\alpha_k(t)>
\nen \\
= &\cos(\theta/2) \left( c_+ |+> +c_- |->\right)
\\
  &~~~~~+i\sin(\theta/2) (c_+ |->_{\sigma_3}|\alpha_k(t)> - c_- |+>_{\sigma_3}|-\alpha_k(t)>
\nen \ea
The first term is still a simple sum of eigenvectors of the
Hamiltonian after the interaction. The second term, however, is
 not. We thus need to follow the evolution of the two states 
$|->_{\sigma_3}|\alpha_k(t_0)>$ and $|+>_{\sigma_3}|-\alpha_k(t_0)>$.
 Since $\sigma_3$ is a constant of the motion after the interaction
again, the evolution takes place completely in the field sector.
Let us look at the first state first. (The evolution of the second
can be derived easily from that for the first because of the
symmetry of the problem.)

I will again work in the Heisenberg representation.
The field obeys
\ba
\dot\phi_-(t,x) = \pi_-(t,x) +\epsilon(t) h(x)
\\
\dot \pi_-(t,x) = \partial^2_x \phi_-(t,x) -m^2 \phi_-(t,x)
\ea
with solution
At the time $t_0$ the field is in the coherent state $|\alpha_k>$.
This can be represented by taking the field operator to be of the form
\ba
\phi_-(t_0,x) = \phi_0(t_0,x)
\\
\pi_-(t_0,x) = \dot\phi_0(t_0,x) +\epsilon(t_0) h(x)
\ea
whee the state $|\alpha_k>$ is the vacuum state for the free
 field $\phi_0$..
We can now solve the equations of motion for $\phi_-$ and obtain
 (again assuming that $\epsilon(t)$ is slowly varying)
\ba
\phi_-(t,x) = \phi_0(t,x) + 2\psi(t,x) \epsilon(t_0)
\\
\pi_-(t,x) = \dot\phi_0(t,x) +2\psi(t,x)\epsilon(t_0) - \epsilon(t) h(x)
\ea
where $\psi(t_0,x)=0$ and $\dot\psi(t_0,x)= h(x)$. Thus again, the field is
 in a coherent state set by both $2\epsilon(t_0) \psi$ and $\epsilon(t)h(x)$.
The field $\psi$ propagates away from the interaction region determined by 
$h(x)$, and I will assume that I am interested in times $t$ a long time after 
the time $t_0$. At these times I will assume that $\int h(x)\psi(t,x) dx=0$.
(This overlap dies out as $1/\sqrt{mt}$. The calculations can be carried 
out for times nearer $t_0$ as well--- the
expressions are just messier and not particularly informative.)

Let me define the new coherent state as $|-\alpha_k(t) +\beta_k(t)>$, where
$\alpha_k$ is as before and
\be
\beta_k(t) = 2\epsilon(t_0)\omega_k{ \tilde\psi}(t,k)
= 2i\epsilon(t_0)  e^{i\omega_k t} \th(k)/ \omega_k
\ee
(The assumption regarding the overlap of $h(x)$
and $\psi(t)$ corresponds to the assumption that
 $\int \alpha_k^*(t)\beta_k(t) dk=0$). Thus the state
 $|->_{\sigma_3} |\alpha_k>$ evolves to the state
$|->_{\sigma_3} |-\alpha_k +\beta_k(t)>$. Similarly, the state
$|+>_{\sigma_3}|-\alpha_k>$ evolves to $|+>_{\sigma_3} |\alpha_k-\beta_k(t)>$.

We now calculate the overlaps of the various states of interest.

\ba
<\alpha_k|\alpha_k\pm \beta_k>
 = <-\alpha_k|-\alpha_k\pm\beta_k>= e^{-\int|\beta_k|^2dk}= J(t_0)
\\
<-\alpha_k|\alpha_k\pm\beta_k> = <\alpha_k|-\alpha_k \pm \beta_k> = J(t)J(t_0)
\\
<-\alpha_k+\beta_k|\alpha_k-\beta_k>
=<-\alpha_k-\beta_k|\alpha_k+\beta_k>=J(t)J(t_0)^4
\ea
The density matrix becomes
\ba
\rho_3= \cos(\theta) \rho_{03} +\sin(\theta)J(t_0) \rho_{02}
\\
\rho_1= J(t) \left( \cos(\theta)+J^4(t_0)\sin(\theta)\right)\rho_{01}
\\
\rho_2(t)= J(t) \left( -\sin(\theta) \rho_{03}
+ (\cos(\theta/2)-J^4(t_0)\sin(\theta))\rho_{02} \right)
\ea
where
\ba
\rho_{03} = \half(|c_+|^2-|c_-|^2)
\\
\rho_{01} = Re ( c_+ c^*_-)
\\
\rho_{02}= Im (c_+ c^*_-)
\ea
If we now let $\epsilon(t)$   go slowly to zero again
 ( to find  the `real' loss of coherence), we find that unless
 $\rho_{01}=\rho_{02}=0$  the system has really lost coherence
during the sudden transition. The maximum real loss of coherence
 occurs if the rotation is a spin flip ($\theta=\pi$) and
$\rho_{03}$ was zero. In that case the density vector dropped to
 $J(t_0)^4$ of its original value. If the density matrix was in
 an eigenstate of $\sigma_3$ on the other hand, the density matrix
remained a coherent density matrix, but the environment was still
 excited by the spin.

We can use the models of a fast or a slow spin flip interaction to discuss
the problem of the tunneling time. As Leggett et al argue\cite{garg}, the spin
system is a good model for the consideration of the behaviour of
 a particle in two wells, with a tunneling barrier between
 the two wells. One view of the transition from
one well to the other is that the particle sits in one well
 for a long time. Then at some random time it suddenly jumps
through the barrier to the other side. An alternative view
would be to see the particle as if it were a
fluid, with a narrow pipe connecting it to the other well-
the fluid slowly sloshing between the two wells.  The former
is supported by the fact that if one periodically  observes
 which of the two wells the particle is in, one sees it
 staying in one well for a long time, and then between
two observations, suddenly finding it in the other well.
 This would, if one regarded it as a classical
particle imply that the whole tunneling must have occurred
 between the two observations. It is as if the system were
in an eigenstate and at some random time an interaction
flipped the  particle from one well to the other. However,
 this is not a good picture. The environment is continually observing the system.
It it really moved rapidly from one to the other, the
 environment would see the
rapid change, and would radiate.  Instead, left on its
 own, the environment in this problem ( with a mass much
 greater than the frequency of transition of the
system) simply adjust continually to the changes in the
system. The tunneling thus
seems to take place continually and slowly.

 \section{Discussion}

The high frequency modes of the environment lead to a loss of coherence
(decay of the off-diagonal terms in the density matrix)
of the system,
but as long as the changes in the system are slow enough this decoherence is false--
 it does not prevent 
the quantum interference of the system. The reason is that the changes 
in the environment caused by these 
modes are essentially tied to the system, they are adiabatic changes
 to the environment which can easily be adiabatically reversed. 
Loosely one can say that coherence is lost by the transfer of information (coherence)
from the system to the environment. However in order for this information to be 
truly lost, it must be carried away by the environment, separated from the
system by some mechanism or another so that it cannot come back into the system.
 In the environment above,
 this occurs when the information travels off to infinity. Thus the loss of
 coherence as represented by the reduced density matrix is in some sense the
 maximum loss of coherence of the system. Rapid changes to the system, or
 rapid decoupling of the system from the environment, will make this a true
 decoherence. However, gradual changes in the system or in the coupling to the 
external world can cause the environment to adiabatically track the system 
and restore the 
coherence apparently lost. 

This is of special importance to understanding the effects of the environmental
cutoff in many environments\cite{garg}. For 
``ohmic"  or ``superohmic" environments ( where $h$ does not fall
 off for large arguments), one has to introduce a cutoff
 into the calculation for the reduced density matrix. This cutoff
 has always been a bit mysterious, especially as the  loss of 
coherence depends sensitively on the value of this cutoff. If one imagines
 the environment to include say the electromagnetic field, what is the right 
value for this cutoff? Choosing the Plank scale seems silly, but what is
 proper value? The arguments of this paper suggest that in 
fact the cutoff is unnecessary except in renormalising the dynamics of the system.
 The behaviour of the environment at
frequencies much higher than the inverse time scale of the system leads 
to a false loss of coherence, a loss of coherence which does not
affect the actual coherence ( ability to interfere with itself) of the system. Thus the 
true coherence is independent of cutoff.

As far as the system itself is concerned, one should regard it as
``dressed" with a polarization of the high frequency components of the 
environment. One should regard not the system itself as 
important for the quantum coherence, but a combination of variables of the 
system plus the environment.What is difficult is the dependence of which 
the degrees of freedom of the environment are simply dressing and which 
are degrees of freedom which can lead to loss of coherence depends 
crucially on the motion and the interactions of the system itself. They are
 history dependent, not simply state dependent. This make it very difficult 
to simply find some transformation which will express the system plus
environment  in terms of variables which are genuinely independent, in 
the sense that if the new variable loose coherence, then that loss is real. 

These observations emphasis the importance of not making too rapid conclusions
 from the decoherence of the system.
This  is especially true in cosmology, where high frequency modes of
the cosmological system are 
used to decohere low frequency quantum modes of the universe. Those high
frequency modes are likely to  behave
adiabatically with respect to the low frequency behaviour of the universe.
 Thus although they will lead to a 
reduced density matrix for the low frequency modes which is apparently
 incoherent, that incoherence is likely to be
a false loss of coherence.

\acknowledgments
I would like to thank the Canadian Institute for Advanced Research 
for their support of this research. This research was carried out 
under an NSERC grant 580441.

\end{document}